\def \ln {{\rm \ ln \  }}

\def \ee {{\rm e}^}

\def \k1 {{1\overk}}

\def \a {\alpha}
\def \b {\beta}
\def \hi {\chi}
\def \l {\lambda}

\def \r {\rho}

\def \k {\kappa }
\def \d {\delta}
\def \o {\omega}

\def \ph{\phi}

\def \m{\mu}
\def \n{\nu}

\def \e#1 {{\rm e}^{#1}}

\def \1p {{1\over  \pi }}
\def \2p {{{1\over  2\pi }}}
\def \4p {{ {1\over 4 \pi }}}
\def \8p {{{1\over 8 \pi }}}

\def \hg {{\hat g}}

\newcommand{\rf}[1]{(\ref{#1})}
\newcommand{\beq}{\begin{equation}}
\newcommand{\eeq}{\end{equation}}
\newcommand{\bea}{\begin{eqnarray}}
\newcommand{\eea}{\end{eqnarray}}
\newcommand{\beas}{\begin{eqnarray*}}
\newcommand{\eeas}{\end{eqnarray*}}

\documentstyle[titlepage,12pt]{article}
\begin{document}

\begin{tabbing}
\` FIT-HE-96-11\\
\` January 1996\\
\` hep-th/xxx
\end{tabbing}
%%%%%%%%%%%%%%%%%%%%%%
\addtolength{\baselineskip}{0.2\baselineskip}
\begin{center}
\vspace{36pt}
  {\large \bf CP$^{N-1}$ model and the Quantized 2d 
                      Black Holes}
\end{center}
\vspace{36pt}
\begin{center}
{\bf Kazuo Ghoroku}\footnote{e-mail: gouroku@dontaku.fit.ac.jp}
\end{center}
\vspace{2pt}
\begin{center}
Department of Physics, Fukuoka Institute of Technology,Wajiro, Higashiku, 
Fukuoka 811-02, Japan
\end{center}
\vspace{36pt}
\begin{center}
{\bf abstract}
\end{center}
We have examined the coupled system of the dilaton gravity and the
$CP^{N-1}$ theory known as a model of the confinement
of massive scalar quarks.
After the quantization of the system, we could see the quantum effect
of the gravitation on the coupling constant of $CP^{N-1}$ model and 
how the coupling constant of dynamically induced gauge field changes
near the black hole configuration.

\newpage
%%%%%%%%%%%%%%%%%%%%%%%%%%%%%%%%  Main  %%%%%%%%%%%%%%%%%%%%%%%%%%%%%%%%%%%%
%%%%%%%%%%%%%%%%%%%%%%%%%%%%%%%%%%%%%%%%%%%%%%%%%%%%%
\section{Introduction}

As for the quantum gravity,
much progress has been made in the 
two dimensional case and it 
becomes being possible to see the quantum effects of
the gravitation on the matter system in terms
of the techniques developed in the string theories. An intersting
discovery is the universal effect ~\cite{poly,gho2,gho1} on the
renormalization group equations of the matter field theory 
which is renormalizable in two dimension. The word
{\it universal} means that the modification caused by the gravitation
depends only on the gravitational model and does not depend of
the models of the matter fields. Among the models of the 2d
gravity, the dilaton gravity is tempting since it
provides the black hole configuration \cite{cal} even if
it is quantized \cite{gho1}.
So, it is possible to examine the effects of the back
ground black holes on the quantities of the matter system.

As for 2d matter field theory, the $CP^{N-1}$ model is
intersting since it has many similar properties to the 4d Yang-Mills
theory; conformal invariance, asymptotic freedom,
dimensional transmutation, instanton configurations and confinement.
So, to study the coupled system of the dilaton gravity and the
$CP^{N-1}$ model
\footnote{In general, $CP^{N-1}$ model is considered for large N, and
it includes many scalar fields. But the string susceptibility of this
coupled system is real because of the characteristic property of the
dilaton gravity \cite{gho1}.}
could give some clue to the effects
of the quantum gravitation and of the black holes
on the Yang-Mills theory. 

In four dimensions, there were the renormalization group analyses
~\cite{tom,park} of matter theories in the curved space and the 
$\beta$-functions were evaluated as a function of the scalar curvature
($R$). The $R$-dependence of the $\beta$-functions has been evaluated
through the one-loop corrections of the matter fields
in a special way, but
the quantum gravity has nothing to do with this calculation. In \cite{park},
the case of the pure Yang-Mills theory
has been examined
in the curved space and the resultant $\beta$-function
shows the asymptotic freedom with respect to $|R|$.

In the two dimensions, there is no physical degree of freedom of gauge
field except for the topological
Coulomb force. Then we can not perform the same calculation with 
ref.\cite{park}
in two dimensions.
So we consider here $CP^{N-1}$ model, which contains many dynamical degrees
of freedom and its properties are similar to 4d Yang-Mills theory,
instead of the 2d Yang-Mills gauge theory in order to see the matter
loop correction in a curved space. On the other hand, we know that
the gauge fields are
dynamically generated in this model, and the long range force mediated
by this gauge field is considered to be
responsible for the confinement of the scalar quarks.
In this sense, we can examine the gravitational effect on the
{\it gauge coupling constant} in the $CP^{N-1}$ model.

The purpose of this paper is to study the $CP^{N-1}$ model coupled with
the dilaton gravity. Quantization of this system is done according
to \cite{gho1} in order to introduce the correct quantum measure.
Since $CP^{N-1}$ model is classically conformal-invariant, this
system seems to be decoupled 
with the gravitation in the conformal gauge. However, we can see that
it couples with the conformal mode through the
dressed factor when the theory is quantized.
Due to this factor, 
we can see the quantum gravitational
effect on the renormalization group equation of the coupling constant
of the $CP^{N-1}$ model.
On the other hand, the gravitational sector
provides the black hole solutions
even if the theory is modified due to the quantization \cite{gho1},
then we can furthermore see how the $CP^{N-1}$ matter 
system is affected by the 
background black hole
configurations. This is performed by the study of the quantized 
effective theory in terms of the 1/N expansion, and it will be shown
that the long range
force of the induced gauge field is largely affected.

\section{$CP^{N-1}$ model}

 The CP$^{N-1}$ model is a kind of a non-linear $\sigma$ model
which can be formulated as the limit of $\lambda\rightarrow 
\infty$ of the following linear theory,
\begin{equation}
 L_{\rm lin}={1 \over 2}{\rm Tr}(\partial_{\mu}\phi\partial^{\mu}\phi)
         -\lambda {\rm Tr}P(\phi).
\end{equation}
where $\phi$ is $N \times N$ traceless Hermitian matrix.
The potential ${\rm Tr}P(\phi)$ is taken so that
its minimum is realized by
\begin{equation}
 \phi= g_0^{-1}\bigl[ N^{{1 \over 2}}zz^{\dagger} -N^{-{1 \over 2}}\it{I}
               \bigr].
\end{equation}
where $z_i$ are N-dimensional complex scalar fields with the constraint,
$$ z^{\dagger} z=1$$ 
and $g_0$ is the coupling constant.
Then the non-linear form of the $CP^{N-1}$ model is written as follows,

\begin{equation}
 L_{\rm nli}= {N \over g_0^2}[\partial_{\mu}z^{\dagger} \partial^{\mu}z
         -g_0^2N^{-1/2}j_{\mu}j^{\nu}], \label{eq:lni}
\end{equation}
where
\begin{equation}
   j_{\mu}={1 \over 2\it{i}}\bigl[ z^{\dagger} \partial_{\mu}z
            - (\partial_{\mu}z^{\dagger})z\bigr].
\label{eq:lag1}
\end{equation}
For the sake of the later convenience, we
rescale $z_i$ as 
$$z_i \rightarrow g_0N^{-1/2}z_i.$$ 
Then, \rf{eq:lni} is rewritten as,
\begin{equation}
 L_{\rm nli}= \partial_{\mu}z\dagger \partial^{\mu}z
         -g_0^2N^{-1/2}j_{\mu}j^{\nu}, \label{eq:ln}
\end{equation}
where $j_{\mu}$ is the same form with \rf{eq:lag1}.
While the constraint is changed as
\begin{equation}
   z^{\dagger} z= N/g_0^2. \label{eq:cons}
\end{equation}

 This model has several intersting properties which
have been cleared through the $1/N$ expansion; (i)The scalar fields $z_i$
acquire their mass, and its scale is characterized by the scale
of the dimensional transmutation which occurs in this theory.
(ii)The second notable point is that the massive scalar fields are 
confined by the long range force of the induced gauge fields. More details
are seen in \cite{witt}.

Since this theory is classically conformal invariant,
it decouples from the gravity for the conformal gauge, $g_{\mu\nu}=
\eta_{\mu\nu}{\rm e}^{2\rho}$. However this is not true
in the quantum theory since the theory would
be corrected through the dressed factor which should be produced by the
quantum gravitational effect as shown below. 

\section{Dilaton gravity}

Here we consider the dilaton gravity
,which couples with $CP^{N-1}$ model, because of the following two 
reasons; (i)The string susceptibility would be real in this coupled
system \cite{gho1}. (ii) The influence of the non-trivial background,
the black hole, on the $CP^{N-1}$ system can be seen through the effective
theory obtained after the quantization. The quantized dilaton gravity
contains the same black hole configuration with the one given by the
classical theory \cite{gho1}.

The action of the dilaton gravity is written as 
\beq
S_{\rm dil}=
{1 \over 4\pi}\int\,d^2z\sqrt{g} e^{-2\phi}
            \bigl[ R+4(\nabla\phi)^2+4\m^2\bigr],  \label{eq:dil}
\eeq
\noindent where $\phi$ represents the dilaton field and
$\nabla_{\mu}$ denotes the covariant derivative with 
respect to the 2d metric $g_{\m\n}$. However we use another form
of action, which is essentially equivalent with \rf{eq:dil} and it
provides the static black hole solutions for non-zero $\mu$, for the
sake of the convenience of our calculations. It is written as 
\cite{gho1},
\begin{equation}
       S_{\rm mod}=-{1 \over 4\pi}\int\,d^2z\sqrt{g}
            \bigl( \hi R+4(\nabla\o)^2
                  + \mu^2[(\l+4)\o^2-\l\hi\bigr),
                                         \label{eq:mod}
\end{equation}
\noindent We can see the equivalency of this model and \rf{eq:dil}
by integrating over the auxiliary fields
$\l$ and $\hi$ and by interpreting
$\o$ as $\exp{(-\ph)}$. So it is trivial to see that 
this action provides the static black hole solutions
as long as the cosmological constant $\m$ is non-zero.
The reason why we choose the action \rf{eq:mod} instead of \rf{eq:dil}
is as follows. Although it is not easy to quantize \rf{eq:dil},
we can make a systematic procedure of quantization in our 
equivalent model \rf{eq:mod}.
Next we consider the coupled system of the $CP^{N-1}$ model
and the dilaton gravity.

\section{Quantized action}

Here we consider the coupled system, 
\begin{equation}
   S=S_{\rm mod}+S_{\rm nli},
\end{equation}
where $S_{\rm nli}=\int d^2z L_{\rm nli}$.
According to \cite{gho2} and \cite{gho1}, the quantization 
with respect to the gravitation is performed. We notice here
that the dynamical modes of $CP^{N-1}$ system are the
$2N-2$ real scalars \cite{cole}. The two redundant freedoms of $z_i$ 
are eliminated as follows; (i) Fix the over all phase of $z_i$ so that
$z_{N}$ is real, then (ii) remove Re$z_{N}$ in terms of
the constraint $z^{\dagger} z=N/g_0^2$. Then we obtain the following
form of $S_{\rm nli}$,
\begin{equation}
  S_{\rm nli}=-{1 \over 8\pi}\int\,d^2z\sqrt{g}
                  g^{\mu\nu}(\delta^{ij}+{g_0^2 \over N}K^{ij})
                  \partial_{\mu}f_i\partial_{\nu}f_j, \label{eq:snli}
\end{equation}
where $z_i\equiv if_i+f_{N+i}$, $i,j=1\sim 2N-2$
\footnote{Here $f_N$ and $f_{2N}$ are lacking since $z_N$ is eliminated.}
and
\bea
      K_{ij}&=&{f_if_j \over 1-
                      g_0^2/N\sum_{i=1}f_i^2}+k_{ij}, \label{eq:Km} \\
      k_{ij}&=&\left\{ \begin{array}{ll}
                   0   & \mbox{for $i,j<N$}     \\
                   f_{i+N-1}f_{j+1-N} & \mbox{for $i<N$, $N\le j$} \\
                   f_{i+1-N}f_{j+N-1} & \mbox{for $j<N$, $N\le i$} \\
                   f_{i+1-N}f_{j+1-N} & \mbox{for $i\le N$, $N\le j$} 
                        \end{array}\right . \label{eq:kij}
\eea

Here we take
the conformal gauge, $g_{\m\n}=\ee{2\r}\hg_{\m\n}$ with a fiducial
metric $\hg_{\m\n}$. 
Due to the metric invariance, the fully quantized action is 
obtained by imposing the conformal invariance with respect to the 
fiducial metric \cite{ddk} on the action. According to this idea, we
firstly obtain the effective action for $\m^2=0$ and $g_0^2/N=0$ limit,
where the system is free from the interactions.
We adopt the following definition of the norms for
each variables, $X^a=(\r, \hi, \o, f_i)$
\footnote{Here, $a=(0,1,2,\cdots,2N-2)$ and the total number of the freedom
is $3+(2N-2)=2N+1$.}

\beq
  \|\d X^a\|^2=\int\,d^2z\sqrt{\hg} (\d X^a)^2.
\label{eq:a4}
\eeq

\noindent Then the quantized effective action could be obtained
in the following form,
\begin{equation}
      S_{\rm eff}^{(0)}={1 \over 4\pi}\int\,d^2z\sqrt{\hg}
            \left[ {1 \over 2}G_{ab}^{(0)}(X)\hg^{\a\b}
              \partial_\a X^a\partial_\b X^b
           +{\hat R}\Phi^{(0)}(X)+T^{(0)}(X) \right], \label{eq:a5}
\end{equation}
and

\beq
   G_{ab}^{(0)}=\pmatrix{\k & -2 &  &  \cr
                         -2 &  0 &  &  \cr
                            &    & -8 &  \cr
                            &    &   & {\bf 1} \cr}, \  \  {}
   \Phi^{(0)}={1 \over 2}\k\r-\hi,
             \ \  T^{(0)}=0, \label{eq:a6}
\eeq
\noindent where ${\bf 1}$ in $G_{ab}^{(0)}$
denotes the unit matrix of dimension $2N-2$ and 
\beq
       \k={26-3-(2N-2) \over 3}={25-2N \over 3}. \label{eq:a7}
\eeq
Here the value of $\kappa$ is the coefficient of the Liouville
action and it is determined by the ghost of the diffeomorphism
gauge fixing and other fields.

The next task is to obtain the quantized action of non-zero
$\mu^2$ and $g_0^2/N$, which are assumed here to be small. There
are two parameters in this theory. For the sake of the simplicity,
we assume here
\begin{equation}
      \mu^2 \cong g_0^2/N \ll 1, 
\end{equation}
and expand the effective action in the series of $\mu^2$ and
$g_0^2/N$.
It is formally written as
\bea
  S_{\rm eff}&=&S_{\rm eff}^{(0)}+\m^2 S_{\rm eff}^{(1a)}
                                 +g_0^2/NS_{\rm eff}^{(1b)} \nonumber \\
             & &\qquad\qquad    + \m^4 S_{\rm eff}^{(2a)}
                           + \m^2g_0^2/N S_{\rm eff}^{(2b)}
                        + (g_0^2/N)^2 S_{\rm eff}^{(2c)}+\cdots  
                                                         \label{eq:eff1} \\
             &=&{1 \over 4\pi}\int\,d^2z\sqrt{\hg}
            \left[ {1 \over 2}G_{ab}(X)\hg^{\a\b}
              \partial_\a X^a\partial_\b X^b
           +{\hat R}\Phi(X)+T(X) \right]. \label{eq:a8}
\eea
where
\bea
G_{ab}&=& G_{ab}^{(0)}+g_0^2/N G_{ab}^{(1)}+\m^4G_{ab}^{(2a)}+\cdots , 
           \nonumber \\
      & & \Phi=\Phi^{(0)}+g_0^2/N\Phi^{(1)}+\m^4\Phi^{(2a)}+\cdots, 
           \nonumber \\
     & & \ \ {} T=T^{(0)} +\m^2 T^{(1)}+\cdots. \label{eq:a9}
\eea
The reason why the terms of order $0(\m^2)$ in $G_{ab}$ and $\Phi$ 
and the term of order $g_0^2/N$ in $T$ do not appear can be seen 
from the eqs.\rf{eq:n4}, \rf{eq:n5} and \rf{eq:n6} given below. 
The second and third terms of eq.\rf{eq:eff1} are given by the procedure
to obtain the so-called dressed factors of the perturbation.

The dressed factors of the terms
being proportional to $\mu^2$ and $g_0^2/N$ in eq.\rf{eq:mod}, and the
higher order terms of $S_{\rm eff}$ are obtained by solving the 
following equations derived from the target space action,

\bea
\nabla^2T-2\nabla\Phi \nabla T&=&{1 \over 2}v'(T),\label{eq:n4} \\
 \nabla^2\Phi-2(\nabla\Phi)^2&=&-{\kappa \over 2}
                       +{1 \over 32}v(T), \label{eq:n5} \\
 R_{ab}-{1 \over 2}G_{ab}R= 
            -2\nabla_{a}\nabla_{b} \Phi &+&G_{ab}\nabla^2\Phi
            +{1 \over 16}\nabla_{a}T\nabla_{b}T-{1 \over 32}G_{ab}
            (\nabla T)^2. \label{eq:n6}
\eea

\noindent where $v(T)=-2T^2+{1 \over 6}T^3+\cdots $ and
$v'=dv/dT$.
$\nabla_{\m}$ denotes the covariant derivative with 
respect to the metric $G_{ab}$. 

Here we estimate $S_{\rm eff}^{(1a,b)}$ in \rf{eq:eff1} in order to 
obtain a simple but a
non-trivial effective action. $S_{\rm eff}^{(1a)}$ is obtained
by solving the lowest order equation of \rf{eq:n4}. And the solution
is the same form with the one given in \cite{gho1}, it is written as
\beq
       S_{\rm eff}^{(1a)}=-{1 \over 4\pi}\int\,d^2z\sqrt{\hg}\ee{2\r}
            \bigl[ (\l+4)(\o^2-{1 \over 4}\r)
                    -\l(\hi-2\r)\bigr].
                                         \label{eq:b5}
\eeq

The third term $S_{\rm eff}^{(1b)}$ can be obtained in a similar way
performed in the non-linear $\sigma$ model \cite{gho2}.
We introduce the dressed factor as,

\begin{equation}
  S_{\rm nli}^{\rm dressed}
          = -{1 \over 4\pi}\int d^2z \sqrt{\hg}
               \exp{(\gamma\rho)}L_{\rm nli}
                                , \label{eq:ln3}
\end{equation}
where $L_{\rm nli}$ is given by \rf{eq:lni}.
Then, rescale $z_i$ as shown above
and parametrize them by $f_i$ as given in eq.\rf{eq:snli},
and we obtain the following expansion of \rf{eq:ln3} with respect to
$g_0^2/N$, 
\begin{equation}
  S_{\rm nli}^{\rm dressed}
          = -{1 \over 4\pi}\int\,d^2z\sqrt{\hg}
             \hg^{\mu\nu}(\delta^{ij}+{g_0^2 \over N}K^{ij}+\delta^{ij}A\rho)
                  \partial_{\mu}f_i\partial_{\nu}f_j
                   +O(({g_0^2 \over N})^2), \label{eq:ln4}
\end{equation}
where we have assumed the following form of expansion for $\gamma$,
\begin{equation}
         \gamma = A{g_0^2 \over N}+O(({g_0^2 \over N})^2), 
                               \label{eq:ln5}
\end{equation}

Then, $S_{\rm eff}^{(1b)}$ can be given by solving the
lowest order ($O(g_0^2/N)$) equations of \rf{eq:n5} and \rf{eq:n6}, in which
$T^{(1)}$ provides the contribution of $O(\mu^4)$ and they can be neglected.
Further we can read the following,
\begin{equation}
  G_{ab}^{(1)} = \left\{ \begin{array}{ll}
                   0   & \mbox{for $a,b\ne i,j$}     \\
         f_if_j+k_{ij}+\delta_{ij}A\rho & \mbox{for $a,b= i,j$} 
                        \end{array}\right. \label{eq:gab}
\end{equation}
from eqs.\rf{eq:Km}, \rf{eq:kij} and \rf{eq:ln4}. 
Then the equations to be solved
are written as follows,

\bea
 \partial^2\Phi^{(1)}-2\partial_0\Phi^{(1)}&=&-{2N-2 \over 4}A, 
                                          \label{eq:n51} \\
 R_{ab}^{(1)}-{1 \over 2}G_{ab}^{(0)}R^{(1)}= 
            -2\partial_a\partial_{b} \Phi^{(1)} &+&
               G_{ab}^{(0)}[\partial^2\Phi^{(1)}
            +{2N-2 \over 4}A]-{1 \over 2}A\delta_a^i\delta_b^j,
             \label{eq:n61}
\eea
where $R^{(1)}_{ab}$ denotes the Ricci curvature of the order 
$O(g_0^2/N)$ and it is written as,
\begin{equation}
 R_{ab}^{(1)}= \left\{ \begin{array}{ll}
                   0   & \mbox{for $a,b\neq i,j$}     \\
                  -(\delta^k_k-2)\delta_{ij} & 
                           \mbox{for $a,b=i,j$} 
                        \end{array}\right.  \label{eq:curv}
\end{equation}
Solving these equations, we obtain the following solutions,
\bea
       \Phi^{(1)}&=&-{1 \over 4}R^{(1)}\rho, 
                                          \label{eq:n52} \\
               A &=& -{2 \over \delta^i_i}R^{(1)}
                  = 4N. \label{eq:n62}
\eea
We notice here that $A$ is positive and this fact is important.

\section{Gravitational effects on $CP^{N-1}$ theory}

From \rf{eq:n62},
we can see the gravitational dressing of the renormalization group
equation of $g_0$, the coupling constant of the $CP^{N-1}$,
according to the procedure of \cite{gho2}. Consider the action
\rf{eq:ln3}, then make the shift of the physical scale,
$\r\rightarrow \r+2dl/\eta$, where $\eta$ 
is defined by the dressed
factor of the cosmological constant in the form $\ee{\eta\r}$. 
Then, this shift is absorbed by the parameters
of the theory and we obtain the following $\beta$-function,
\beq
   \b(g_0)=-{dg_0^2 \over dl}
           = -2N{2 \over \eta}g_0^4.
                        \label{eq:beq}
\eeq
The factor $\eta$ is obtained by solving the lowest order 
of eq.\rf{eq:n4} where 
${\hat T}=\ee{\eta\r}$ \cite{gho2}, and it is given as
$\eta=2/\kappa$. 
The result \rf{eq:beq} is consistent with the previous
analyses ~\cite{gho2,gho1}. This $\beta$-function is
equivalent with the one given in \cite{din} without gravitation 
except for the gravitational dressed factor $2/\eta$.
\footnote{We should notice here that the coupling constant used in
\cite{din} is different from the one using here by the
factor two due to its definition.}
This is the universal feature of the $\beta$-functions which have
been obtained with quantum gravitation.
From \rf{eq:n52}, the shift of the background charge, which is read
from the coefficient of $\rho$ in $\Phi$, is also observed, and it
is consistent with the so called c-theorem \cite{zam}.

Next, we investigate the effect of the background manifold on the
effective $CP^{N-1}$ model. We restrict our attention to the case of
small $\mu^2$. Then the effective action can be approximated by the
following,
\begin{equation}
 S_{\rm eff}=S_{\rm eff}^{(0)}+S_{\rm eff}^{(1a)}
             +S_{\rm nli}^{\rm dressed}. \label{eq:effa}
\end{equation}
The terms of the right hand side are given by \rf{eq:a5}, \rf{eq:b5}
and \rf{eq:ln3} respectively. With respect to the last term, 
$S_{\rm nli}^{\rm dressed}$, we supposed that all the higher order corrections
are included in the factor $\gamma$. This assumption is correct at least
to $O(g_0^2/N)$. 

 The theory given by \rf{eq:effa}
provides a black hole configuration for $z_i=0$, and we obtain the same
form of the curvature with the classical dilaton gravity
as shown in \cite{gho1}. With respect to the $CP^{N-1}$ part,
the instanton solutions exist as
the nontrivial configuration of $z_i$, but we do not consider them here
in order to consider the black hole configuration as a vacuum configuration
of the coupled system.
We do not consider here the both configurations of black hole and the 
instanton.
We here investigate the dynamical properties of $CP^{N-1}$ through the 1/N
expansion 
since the essential features of the $CP^{N-1}$ model have been seen by
the 1/N expansion \cite{witt} rather than the investigation through 
the instantons \cite{din}.
We notice here that
we can see the effect of the background manifold
on the $CP^{N-1}$ part
in the conformal gauge after the quantization of the gravitation
for the first time.
In this sense, this effect can be said as a 
quantum gravitational effect.

We set the fiducial metric as $\hg_{\mu\nu}=\eta_{\mu\nu}$ and
introduce the auxiliary field $A_{\mu}$ as follows,
\begin{eqnarray}
  L_z &=& L_{\rm nli}^{\rm dressed}+{g_0^2 \over N}
           ({\rm e}^{\gamma\rho/2}j_{\mu}
             +{N \over g_0^2}A_{\mu})^2    \\
     &=& {\rm e}^{\gamma\rho}(D_{\mu}z)^{\dagger}D^{\mu}z, 
\label{eq:amu}
\end{eqnarray}
where $D_{\mu}=\partial_{\mu}+i{\rm e}^{-\gamma\rho/2}A_{\mu}$. The 
integration over $A_{\mu}$ leads to the original effective action.
Further, we add the term
\begin{equation}
   -\sigma (z^{\dagger}z-N/g_0^2) \label{eq:con}
\end{equation}
to $L_z$ by using the lagrange multiplier $\sigma$ in order to provide
the constraint for $z_i$. Then the integration over $z_i$ and $z_i^{\dagger}$
has been performed and the following action is induced,
\begin{equation}
    S_{\rm ind}= {i \over 2}N{\rm Trln}
              (D_{\mu}{\rm e}^{\gamma\rho}D_{\mu}z+\sigma)
               +{N \over g^2_0}\int d^2x \sigma. \label{eq:ind}
\end{equation}
This action can be separated into two parts, (i) $A_{\mu}$ independent
and (ii) $A_{\mu}$ dependent terms, and they are denoted by $S_{\sigma}$
and $S_{A}$ respectively, namely
\begin{equation}
    S_{\rm ind}=S_{\sigma}+S_A. \label{eq:ind2}
\end{equation}
These induced actions are estimated here for small $\gamma$, this
assumption is valid for small $g_0^2/N$ since $\gamma$ is expanded
by the power series of $g_0^2/N$ as shown above.

After the calculation of $S_{\sigma}$, we obtain the effective
potential for $\sigma$,
\begin{equation}
 V_{\rm eff}(\sigma)=-N\sigma[{1 \over g_0^2}+
      {1 \over 8\pi}{\rm e}^{\gamma\rho}(\ln{\sigma \over \Lambda^2}-1)],
               \label{eq:sig}
\end{equation}
where $\Lambda$ denotes the cutoff. From this, the renomalized coupling
constant at mass scale $M$ is defined by
\begin{equation}
     {1 \over g^2(M)}= -N^{-1}
         {\partial V_{\rm eff}(\sigma) \over \partial \sigma}\mid
         _{\sigma =M^2},    
               \label{eq:coup}
\end{equation}
and we obtain
\begin{equation}
   g^2(M)={g_0^2 \over 1+
      {1 \over 8\pi}g_0^2{\rm e}^{\gamma\rho}\ln{M^2 \over \Lambda^2}}.
                 \label{eq:cp1}
\end{equation}
This result is different from the running coupling constant, which is
obtained in the flat space-time, by the factor $\exp{(\gamma\rho)}$.
Considering the black hole configuration \cite{gho1}, this factor
can be written in terms of the scalar curvature, $R$, 
and the black hole mass, $M_{\rm bh}$, as follows,
\begin{equation}
     \exp{(\gamma\rho)}=({R \over 4\mu M_{\rm bh}})^{\gamma/2}.
                 \label{eq:cp2}
\end{equation}
Since $R$ diverges at the singular point, just on the black hole, then
the coupling constant approaches to zero near this point. 
This implies the asymptotic freedom
of the $CP^{N-1}$ coupling constant when the system approaches to the
black hole. This property is similar to the 4d Yang-Mills
case where the effective gauge coupling constant $e_{eff}$ is given
\cite{park} as follows with the bare coupling $e_0$,

\begin{equation}
    e_{eff}^2=\frac{e_0^2}{1+Ce_0^2{\rm ln}({R \over R_0})}, \label{eq:pa}
\end{equation}
where $C$ is a constant depending on the gauge group and $R_0$
denotes some scale.
This result shows the asymptotic
freedom of the gauge coupling constant
with respect to the scalar curvature. 
However we should compare the result \rf{eq:pa} with
the effective {\it gauge coupling constant}, which is dynamically induced
in $CP^{N-1}$ model,
rather than $g_0$, because the induced gauge interaction
is responsible for the confinement of the scalar quark $z_i$.

This point can be examined by calculating the $A_{\mu}$-dependent part
$S_A$. We perform the calculation
of this action after replacing $\sigma$ by its extremal point,
$\sigma_0$, which is
obtained from $V_{\rm eff}$ given in \rf{eq:sig} as,
\begin{equation}
   {\partial V_{\rm eff} \over \partial \sigma}
       \mid_{\sigma=\sigma_0}=0.
                 \label{eq:cp3}
\end{equation}
and
\begin{equation}
  \sigma_0=M^2{\rm exp}[-{8\pi \over g_0^2}{\rm e}^{-\gamma\rho}].
                 \label{eq:cp4}
\end{equation}
This is a dimensional transmutation. The dimensionless coupling constant
$g_0$ has been replaced by the dimensionful parameter $\sigma_0$, which
is invariant under the renormalization group transformation. 

$S_A$ is expanded by $\gamma$ as follows,
\begin{equation}
    S_A= {i \over 2}N[{\rm Tr}f(A) -\gamma{\rm Tr}(\rho f(A))
                 +O(\gamma^2)],
                 \label{eq:cp5}
\end{equation}
where
\begin{equation}
    f(A)= -{1\over \partial^2+\sigma_0}A_{\mu}A^{\mu}
         +{1\over 2}{1\over \partial^2+\sigma_0}
           (\partial_{\mu}A^{\mu}+A_{\mu}\partial^{\mu})
          {1\over \partial^2+\sigma_0}
           (\partial_{\nu}A^{\nu}+A_{\nu}\partial^{\nu}).
                 \label{eq:cp6}
\end{equation}
Using the approximation, $\exp{(-\gamma\rho)}\sim 1-\gamma\rho$,
and taking the low momentum limit 
\footnote{This limit is valid since we are investigating 
the long range force of the induced gauge theory.}
of the inner propagators
of the scalar fields $z_i$ \cite{cole}, the following
result is obtained,
\begin{equation}
    S_A= -{N \over 48\pi\sigma_0^2}\int d^2x
                 {\rm e}^{-\gamma\rho}
           (\partial_{\mu}A_{\nu}-\partial_{\nu}A_{\mu})^2.
                 \label{eq:cp7}
\end{equation}
For $\gamma=0$, this is equivalent to the induced kinetic term of gauge fields
which is given in \cite{cole}, and the effective coupling constant of the 
long range force is identified with $\sigma_0$. We can see that
the dressed factor $\exp{(-\gamma\rho)}$ appears in $S_A$
and the effective gauge-coupling constant
is changed to
\begin{equation}
    g_{\rm eff}^2 = \sigma_0^2 \exp{(\gamma\rho)}
                          \label{eq:cp8}
\end{equation}
when the gravitation is switched on.
Near the singularity of the black hole, $\sigma_0$ approaches to the
constant $M^2$ as seen from \rf{eq:cp4} and \rf{eq:cp2}, then 
$g_{\rm eff}^2$ diverges with the increasing curvature. This fact
implies that the massive charged particles $z_i$ are very strongly
tied by the long range force.
This result seems to be in contradiction to the case of the 4d
Yang-Mills case of \cite{park}.

\section{Concluding Remark}

$CP^{N-1}$ model has been studied by combining it with the dilaton
gravity which provides the black hole configuration. The quantum
effect of the gravitation produces the dressed factor for the 
$CP^{N-1}$ part, and we can see the gravitational effects on the
coupling constant of the model due to this factor. The resultant
renormalization group equation receives the common correction to the
case of other interacting matter fields, Sine-Goldon and O(N) 
non-linear $\sigma$ model. Further, we could see the effect of the
black hole configurations on the long range force which is dynamically
induced in $CP^{N-1}$ model and it is responsible for the confinement
of the massive scalar quark. Near the black hole singularity, this
force becomes very strong. This result seems to be curious since
the gauge coupling constant shows asymptotic freedom with the magnitude
of the scalar curvature in four dimension.

%%%%%%%%%%%%%%%%%%%%%%%%%%%  Ref %%%%%%%%%%%%%%%%%%%%%%%%%%%%%%%%%%%%%%%%%

\end{document}